\documentclass[a4paper]{easychair} 

\usepackage{xspace}
\usepackage{amssymb,amsmath}
\usepackage[utf8]{inputenc}
\usepackage{ifthen}
\usepackage{microtype}
\usepackage{hyperref}
\hypersetup{
  colorlinks=true, %
  citecolor=blue, %
  urlcolor=blue, %
  linkcolor=blue, %
}

\theoremstyle{plain}
\newtheorem{theorem}{Theorem}

\newtheorem{lemma}[theorem]{Lemma}

\theoremstyle{definition}
\newtheorem{definition}[theorem]{Definition}

\newcommand\mc[1]{\ensuremath{\mathcal{#1}}\xspace}
\newcommand\msf[1]{\ensuremath{\mathsf{#1}}\xspace}
\def\systemname#1{\text{\msf{#1}}\xspace}
\newcommand{\concon}{\systemname{ConCon}}
\newcommand\RR{\mc{R}}
\newcommand\VV{\mc{V}}

\newcommand{\Var}{\ensuremath{\VV\textsf{\textup{ar}}}\xspace}
\newcommand\RRu{\RR_{\msf{u}}}
\newcommand\etal{et\ al{.}}

\newcommand\thmref[1]{Theorem~\ref{thm:#1}}
\newcommand\togreek[1]{\ifcase#1\relax\or\alpha\or\beta\else\fi}
\newcommand\rel[1]{\xr[\togreek{#1}]{}}
\newcommand\ler[1]{\xl[\togreek{#1}]{}}
\newcommand\isafor{\textsf{Isa\kern-0.2exF\kern-0.2exo\kern-0.2exR}\xspace}
\newcommand\ceta{\textsf{C\kern-0.2exe\kern-0.5exT\kern-0.5exA}\xspace}
\newcommand\relcomp{\cdot}
\newcommand\IF{\Leftarrow}
\newcommand\crule[1]{\ell_{#1} \to r_{#1} \IF c_{#1}}
\newcommand\epar[2][]{\xphr[#1]{#2}}
\newcommand\eparl[2][]{\xphl[#1]{#2}}
\newcommand\rstep[2][\RR]{\xr[#1]{#2}}
\newcommand\cstepn[2][]{\xr[#1]{#2}}
\newcommand\cstepnl[2][]{\xl[#1]{#2}}
\newcommand\cstep[2][\RR]{\xr[#1]{#2}}
\newcommand\seq[2][n]{#2_1,\ldots,#2_{#1}}
\newcommand\isaforchangeset{a2cd778de34a}
\newcommand\isaforpath{http://cl2-informatik.uibk.ac.at/rewriting/%
  mercurial.cgi/IsaFoR/file/\isaforchangeset/IsaFoR}
\newcommand\thyref[1]{\href{\isaforpath/#1.thy}{\nolinkurl{#1}}}
\newcommand\factref[3]{\href{\isaforpath/#2.thy\#l#3}{\nolinkurl{#1}}}

\newcommand\MnFont[1]{
 \DeclareFontFamily{U}{MnSymbol#1}{}
 \DeclareSymbolFont{MnSy#1}{U}{MnSymbol#1}{m}{n}
 \SetSymbolFont{MnSy#1}{bold}{U}{MnSymbol#1}{b}{n}
 \DeclareFontShape{U}{MnSymbol#1}{m}{n}{
     <-6>  MnSymbol#15
    <6-7>  MnSymbol#16
    <7-8>  MnSymbol#17
    <8-9>  MnSymbol#18
    <9-10> MnSymbol#19
   <10-12> MnSymbol#110
   <12->   MnSymbol#112}{}
 \DeclareFontShape{U}{MnSymbol#1}{b}{n}{
     <-6>  MnSymbol#1-Bold5
    <6-7>  MnSymbol#1-Bold6
    <7-8>  MnSymbol#1-Bold7
    <8-9>  MnSymbol#1-Bold8
    <9-10> MnSymbol#1-Bold9
   <10-12> MnSymbol#1-Bold10
   <12->   MnSymbol#1-Bold12}{}
}
\MnFont{A}\MnFont{B}\MnFont{C}\MnFont{D}
\DeclareMathSymbol{\mnshortparallel}{\mathbin}{MnSyC}{'376}

\newcommand{\overlayrel}[4]{\mathrel{%
 \def\next##1##2{%
  \setbox0=\hbox{$##1#3$}%
  \setbox1=\hbox to\wd0{$##1\hfil\mkern#1mu#4\mkern#2mu\hfil$}%
  \dp1=\dp0\ht1=\ht0\wd0=0pt\box0\box1%
 }%
 \mathpalette\next{}%
}}
\newcommand\smallparallel{\mnshortparallel}

\newcommand{\xprightarrow}[2][]%
 {\overlayrel02{\xrightarrow[#1]{#2}}\smallparallel}
\newcommand{\xpleftarrow}[2][]%
 {\overlayrel20{\xleftarrow[#1]{#2}}\smallparallel}
\newcommand{\phrightarrow}{\overlayrel02\hookrightarrow\smallparallel}
\newcommand{\phleftarrow}{\overlayrel20\hookleftarrow\smallparallel}
\newcommand{\xphrightarrow}[2][]%
 {\overlayrel02{\xhookrightarrow[#1]{#2}}\smallparallel}
\newcommand{\xphleftarrow}[2][]%
 {\overlayrel20{\xhookleftarrow[#1]{#2}}\smallparallel}

\makeatletter

\newcommand{\r@rrow}[3]{%
  \newcommand{#1}[2][]{%
    \def\next{#2\@ifempty{##1}{}{_{##1}}\@ifempty{##2}{}{^{##2}}}%
    \mathchoice{#3[##1]{##2}}{\next}{\next}{\next}%
  }%
}
\newcommand{\l@rrow}[3]{%
  \newcommand{#1}[2][]{%
    \def\next####1{%
      \setbox0=\hbox{$####1\vphantom{#2}\@ifempty{##1}{}{_{\vphantom{##1}}}%
      \@ifempty{##2}{}{^{##2}}$}%
      \setbox1=\hbox{$####1\vphantom{#2}\@ifempty{##1}{}{_{##1}}%
      \@ifempty{##2}{}{^{\vphantom{##2}}}$}%
      \setbox2=\vbox{\hbox to\wd0{}\hbox to\wd1{}}%
      \mathrel{\hskip\wd2\hskip-\wd0\box0\hskip-\wd1\box1{#2}}%
    }%
    \mathchoice{#3[##1]{##2}}{\next\textstyle}%
    {\next\scriptstyle}{\next\scriptscriptstyle}%
  }%
}

\l@rrow{\xl}{\leftarrow}{\xleftarrow}
\r@rrow{\xr}{\rightarrow}{\xrightarrow}
\l@rrow{\xphl}{\phleftarrow}{\xphleftarrow}
\r@rrow{\xphr}{\phrightarrow}{\xphrightarrow}

\makeatother 

\begin{document}

\title{%
Level-Confluence of 3-CTRSs in Isabelle/HOL%
\thanks{The research described in this paper is supported by FWF (Austrian
Science Fund) project P27502.}}

\titlerunning{Level-Confluence of 3-CTRSs in Isabelle/HOL}

\author{Christian Sternagel \and Thomas Sternagel}

\institute{University of Innsbruck, Innsbruck, Austria \\
\email{\{christian.sternagel, thomas.sternagel\}@uibk.ac.at}}

\authorrunning{Sternagel and Sternagel}
\indexedauthor{Sternagel, Christian}
\indexedauthor{Sternagel, Thomas}

\maketitle

\begin{abstract}
We present an Isabelle/HOL formalization of an earlier result by Suzuki,
Middeldorp, and Ida; namely that a certain class of conditional rewrite systems
is level-confluent. Our formalization is basically along the lines of the
original proof, from which we deviate mostly in the level of detail as well as
concerning some basic definitions.
\end{abstract}

\section{Introduction}

In the realm of standard term rewriting, many properties of term rewrite systems
(TRSs) can be conveniently checked ``at the push of a button'' due to a wealth
of existing automated tools. To maximize the reliability
of this approach, such automated tools are progressively complemented by
certifiers, that is, verified programs that rigorously ensure that the output of an automated
tool for a given input is correct. At the time of writing the prevalent
methodology for certifier development consists of the following two phases:
First, employ a proof assistant (in our case Isabelle/HOL~\cite{Isabelle})
in order to formalize the underlying theory,
resulting in a \emph{formal library} (in our case
\isafor,\footnote{\url{http://cl-informatik.uibk.ac.at/software/ceta/}} an
\emph{Isabelle/HOL Formalization of Rewriting}).
Then, verify a program using this library,
resulting in the actual certifier (in our case \ceta~\cite{ceta}).

Our ultimate goal is to establish the same state of the art also for conditional
term rewrite systems (CTRSs). As a starting point, we present our Isabelle/HOL
formalization of the following result:
\begin{lemma}[{Suzuki \etal~\cite[Corollary 4.7]{SMI95}}]
\label{lem:main}
Orthogonal, properly oriented, right-stable 3-CTRSs are level-confluent.\qed
\end{lemma}
Which is actually a corollary of a more general result, whose statement --
together with a high-level overview of its proof -- we defer until after we have
established some necessary preliminaries.

The development we describe in this note is now part of the \isafor
library and is freely available for inspection at:
\begin{quote}\small
\url{\isaforpath/Conditional_Rewriting/Level_Confluence.thy}
\end{quote}
Throughout the remainder we will from time to time refer to the Isabelle/HOL
sources of our development (by active hyperlink).

\section{Preliminaries}

We assume familiarity with (conditional) term rewriting \cite{BN98,O02}.
In the sequel we consider oriented 3-CTRSs where extra variables in conditions
and right-hand sides of rewrite rules are allowed,
i.e., for all rules $\crule{}$ in the CTRS we only demand
${\Var(r)} \subseteq {\Var(\ell, c)}$.
For such systems extended TRSs $\RR_n$ are inductively defined for each
level $n \geqslant 0$ as follows
\begin{align*}
  \RR_0 &= \varnothing \\
  \RR_{n+1} &= \{ \ell\sigma \to r\sigma \mid \crule{} \in \RR
  \text{ and $s\sigma \rstep[\RR_n]{*} t\sigma$ for all $s \approx t \in c$}
  \}
\end{align*}
where $\rstep[\RR_n]{}$ denotes the standard (unconditional) rewrite relation of
the TRS $\RR_n$.
We write $s \cstep[\RR]{} t$ if we have $s \rstep[\RR_n]{} t$ for some $n \geqslant
0$.
Moreover, for brevity, the latter is written $s \cstepn[n]{} t$ whenever the
corresponding CTRS is clear from the context.
Given two variable disjoint variants $\crule1$ and $\crule2$ of rules in a
CTRS $\RR$, a function position $p$ in $\ell_1$, and a most general unifier
(mgu) $\mu$ of
$\ell_1|_p$ and $\ell_2$; we call
the triple $(\crule1, \crule2, p)$ a \emph{conditional overlap} of $\RR$.
A conditional overlap $(\crule1, \crule2, p)$ with mgu $\mu$ is
\emph{infeasible} (that is, cannot occur during actual rewriting) if there is no
substitution $\sigma$ such that $s\sigma \cstep{*} t\sigma$ for all
$s \approx t$ in $c_1\mu,c_2\mu$.
%

\smallskip
\textsl{A note on permutations.}
At the highly formal level of Isabelle/HOL (which we tend to avoid in the
following exposition) we employ an existing formalization of \emph{permutation
types} (that is, types that contain variables which may be renamed w.r.t.\ a
given permutation) to tackle variable renamings, renaming rules apart, and
checking whether two rules are variants of each other. This abstract view on renamings (as opposed
to explicit renaming functions on strings) proved to be useful in previous
applications~\cite{HMS14,HMS14a}.
\smallskip

We call a CTRS \emph{almost orthogonal}~\cite{H95} \emph{(modulo infeasibility)}
if it is left-linear and all its conditional overlaps are either infeasible or
take place at root position ($\ell_1\mu = \ell_2\mu$) and are either between
variants of the same rule or also result in syntactically equal right-hand sides
($r_1\mu = r_2\mu$).
A CTRS $\RR$ is called \emph{properly oriented} if for all rules $\ell \to r
\IF s_1 \approx t_1, \ldots, s_k \approx t_k \in \RR$ where $\Var(r)
\not\subseteq\Var(\ell)$ and $1 \leqslant i \leqslant k$ we have
$\Var(s_i)\subseteq\Var(\ell, t_1, \ldots, t_{i-1})$.
It is called \emph{right-stable} if for every rule $\ell \to r
\IF s_1 \approx t_1, \ldots, s_k \approx t_k \in \RR$ and
$1 \leqslant i \leqslant k$ we have
$\Var(\ell, s_1, \ldots, s_i, t_1, \ldots, t_{i-1}) \cap \Var(t_i) = \varnothing$
and $t_i$ is either a linear constructor term or a ground $\RRu$-normal form.

We say that two binary relations $\rel1$ and $\rel2$ have
the \emph{commuting diamond property}~\cite{BN98}, whenever
${\ler1} \relcomp {\rel2} \subseteq {\rel2} \relcomp {\ler1}$.
Moreover, we adopt the notion of \emph{extended parallel rewriting} from Suzuki
\etal~\cite{SMI95}.
\begin{definition}
Let $\RR$ be a CTRS.  We say that there is an \emph{extended parallel
$\RR$-rewrite step at level $n$} from $s$ to $t$, written $s \epar[\RR_n]{} t$
(or $s \epar[n]{} t$ for brevity),
whenever we have a multihole context $C$, and sequences of terms $\seq[k]{s}$
and $\seq[k]{t}$, such that $s = C[\seq[k]{s}]$, $t = C[\seq[k]{t}]$, and for
all $1\leqslant i \leqslant k$ we have one of $(s_i, t_i) \in \RR_n$ (that is, a root-step at
level $n$) and $s_i \cstepn[n-1]{*} t_i$.
\end{definition}
Suzuki \etal~\cite{SMI95}, state this definition slightly differently, that is, instead of
multihole contexts they try to rely exclusively on sets of positions:
\begin{quote}\sl
We write $s \epar[n]{} t$ if there exists a subset $P$ of pairwise disjoint positions in $s$
such that for all $p \in P$ either
$(s|_p, t|_p) \in \RR_n$ or $s|_p \cstepn[n-1]{*} t|_p$.
\end{quote}
While it is quite clear what is meant, a slight problem (at least for a formal
development inside a proof assistant) is the fact that this definition does not
enforce $t$ to be exactly the same as $s$ outside of the positions in $P$, that
is, it does not require the multihole context around the $|P|$ rewrite
sequences to stay the same. In order to express this properly, it seems
unavoidable to employ multihole contexts (or something equivalent).

In the remainder we employ the convention that the number of holes of a
multihole context, is denoted by the corresponding lower-case letter, e.g., $c$
for a multihole context $C$, $d$ for $D$, $e$ for $E$, etc.

\section{The Main Result}

As remarked in the last two sections of Suzuki \etal~\cite{SMI95}, we actually
consider \emph{almost orthogonal} systems modulo \emph{infeasibility}.
We are now in a position to state the main theorem.
\begin{theorem}[{Suzuki \etal~\cite[Theorem 4.6]{SMI95}}]
\label{thm:main}
Let $\RR$ be an almost orthogonal (modulo infeasibility), properly oriented,
right-stable 3-CTRS. Then, for
any two levels $m$ and $n$, the extended parallel rewrite relations 
$\epar[m]{}$ and $\epar[n]{}$, have the
commuting diamond property.
\end{theorem}

As a special case of the above theorem, we obtain that for a fixed level $n$,
the relation $\epar[n]{}$ has the diamond property. Moreover, it is well
known that whenever a relation $S$ with the diamond property, is between a
relation $R$ and its reflexive, transitive closure (that is, $R \subseteq S
\subseteq R^*$), then $R$ is confluent. Taken together, this yields
level-confluence of $\cstep[\RR]{}$, since clearly
${\cstepn[n]{}} \subseteq {\epar[n]{}} \subseteq {\cstepn[n]{*}}$.

We now give a high-level overview of the proof of \thmref{main}.
The general structure is similar to the one followed by Suzuki
\etal~\cite{SMI95}, only that we employ multihole contexts instead of sets of
positions. Therefore, we do not give all the details (if you are interested, see
\thyref{Conditional_Rewriting/Level_Confluence}, starting from
\factref{comm_epar_n}{Conditional_Rewriting/Level_Confluence}{1499} in line 1499), but mostly
comment on the parts that differ (if only slightly).
\begin{proof}[Proof (Sketch) of \thmref{main}]
We proceed by complete induction on $m + n$. By induction hypothesis (IH) we may
assume the result for all $m' + n' < m + n$. Now consider the peak
$t \eparl[m]{} s \epar[n]{} u$. If any of $m$ and $n$ equals $0$, we are
done (since ${\epar[0]{}}$ is the identity relation). Thus we may assume $m
= m' + 1$ and $n = n' + 1$ for some $m'$ and $n'$. By the definition of extended
parallel rewriting, we obtain
multihole contexts $C$ and $D$, and sequences of terms $\seq[c]{s}$,
$\seq[c]{t}$, $\seq[d]{u}$, $\seq[d]{v}$, such that
$s = C[\seq[c]{s}]$ and $t = C[\seq[c]{t}]$,
as well as
$s = D[\seq[d]{u}]$ and $u = D[\seq[d]{v}]$;
and
$(s_i, t_i) \in \RR_m$ or $s_i \cstepn[m']{*} t_i$ for all $1 \leqslant i
\leqslant c$,
as well as
$(u_i, v_i) \in \RR_n$ or $u_i \cstepn[n']{*} v_i$ for all $1 \leqslant i
\leqslant d$.

Now we identify the common part $E$ of $C$ and $D$,
employing the semi-lattice properties of multihole contexts (see
\thyref{Rewriting/Multihole_Context}), that is, $E = C
\sqcap D$. Then $C = E[\seq[e]{C}]$ and $D = E[\seq[e]{D}]$ for some multihole
contexts $\seq[e]{C}$ and $\seq[e]{D}$ such that for each $1 \leqslant i
\leqslant e$ we
have $C_i = \Box$ or $D_i = \Box$. This also means that there is a sequence of
terms $\seq[e]{s'}$ such that $s = E[\seq[e]{s'}]$ and for all $1 \leqslant i
\leqslant
e$, we have $s'_i = C_i[s_{k_i},\ldots,s_{k_i + c_i - 1}]$ for some
subsequence $s_{k_i},\ldots,s_{k_i + c_i - 1}$ of $\seq[c]{s}$ (we denote similar
terms for $t$, $u$, and $v$ by $t'_i$, $u'_i$, and $v'_i$, respectively).
Moreover, note that by construction $s'_i = u'_i$ for all $1
\leqslant i \leqslant e$. Since extended parallel rewriting is closed under multihole
contexts (see
\factref{epar_n_mctxt}{Conditional_Rewriting/Level_Confluence}{497}), it
suffices to show that for each $1 \leqslant i \leqslant e$ there is
a term $v$ such that $t'_i \epar[n]{} v \eparl[m]{} v'_i$, in order to conclude the
proof. We concentrate on the case that $C_i = \Box$ (the case $D_i = \Box$
is completely symmetric). Moreover,
note that when we have $s'_i \cstepn[m']{*} t'_i$, the proof concludes by
IH (together with some basic properties of the involved relations), and
thus we remain with $(s'_i, t'_i) \in \RR_m$.
At this
point we
distinguish the following cases:
\begin{enumerate}
\item ($D_i = \Box$).
Also here, the non-root case $u'_i \cstepn[n']{*} v'_i$ is covered by the IH. Thus, we may
restrict to $(u'_i, v'_i) \in \RR_n$, giving rise to a root overlap. Since $\RR$
is almost orthogonal (modulo infeasibility), this means that either the
resulting conditions are not satisfiable or the resulting terms are the same (in
both of these cases we are done) or two variable disjoint variants of the same
rule
$\crule{}$ were involved. Without extra variables in $r$, this is the end of the
story; but since we also want to cover the case where $\Var(r) \not\subseteq
\Var(l)$, we have to reason why this does not cause any trouble. This case is
finished by a technical lemma
(see \factref{trs_n_peak}{Conditional_Rewriting/Level_Confluence}{985})
that shows, by induction on the number of
conditions in $c$, that we can join the two respective instances of the
right-hand side $r$ by extended
parallel rewriting. (This is also where proper orientedness and right-stability
of $\RR$ is first used, that is, were we to relax this properties, we had to
adapt the technical lemma.)

\item ($D_i \neq \Box$).
Then for some $1 \leqslant k \leqslant d$, we have
$(u_j, v_j) \in \RR_n$ or $u_j \cstepn[n']{*} v_j$ for all $k \leqslant j
\leqslant k + d_i
- 1$, that is, an extended parallel rewrite step of level $n$ from
$s'_i = u'_i = D_i[u_{k_i},\ldots,u_{k_i+d_i-1}]$ to $D_i[v_{k_i},\ldots,v_{k_i+d_i-1}] =
v'_i$. Since $\RR$ is almost orthogonal (modulo infeasibility) and, by $D_i \neq
\Box$, root overlaps are excluded,
the constituent parts of the extended parallel step from $s'_i$ to $v'_i$ take place
exclusively inside the substitution of the root-step to the left (which is
somewhat obvious -- as also stated by Suzuki \etal~\cite{SMI95} -- but
surprisingly hard to formalize, see
\factref{epar_n_varpeak'}{Conditional_Rewriting/Level_Confluence}{741}, even
more so when having to deal with infeasibility).
We again close this case by induction on the number of conditions making use of
proper orientedness and right-stability of $\RR$,
see \factref{epar_n_varpeak}{Conditional_Rewriting/Level_Confluence}{1319}
for details.
\qedhere
\end{enumerate}
\end{proof}

\section{Conclusions and Future Work}

In the original paper~\cite{SMI95} the proof of \thmref{main} 
begins after
only three definitions (proper orientedness, right-stability, and extended
parallel rewriting) and stretches across two pages, including two figures.

By contrast, in our formalization we need 8 definitions and 42 lemmas (mainly
stating properties of extended parallel rewriting) before we can start 
with the main proof. Furthermore, we need two auxiliary technical lemmas to cover the induction proofs
on the number of conditions which are nested inside the main case analysis. All
in all, resulting in a theory file of about 1500 lines.
This yields a \emph{de Bruijn factor} of approximately 18, that is, for every
line in the original ``paper proof,'' our formal proof development contains 18
lines of Isar (the formal language of Isabelle/HOL).

In the latest version of our formalization we further relaxed the condition for
conditional overlaps to be infeasible (making the result
applicable to a larger class of systems) and proved that the main result still
holds. More concretely, a conditional overlap $(\crule1, \crule2, p)$ with mgu
$\mu$ is infeasible iff
\[
\forall m\,n.\, {\cstepnl[m]{*}\cdot\cstepn[n]{*}} \subseteq
{\cstepn[n]{*}\cdot\cstepnl[m]{*}}
\implies
\nexists \sigma.\,
  (\forall s \approx t \in c_1\mu.\, s\sigma \cstepn[m]{*} t\sigma) \land
  (\forall s \approx t \in c_2\mu.\, s\sigma \cstepn[n]{*} t\sigma).
\]
That is, we may assume ``level-commutation'' (which is called shallow-confluence
in the literature) when showing that the combined conditions of two rules are
not satisfiable. This may be helpful, since it allows us to turn diverging sequences
(as would for example result from two conditions with identical left-hand sides) into
joining sequences.

\paragraph{Future Work.}
The ultimate goal of this formalization is of course to certify level-confluence
proofs of conditional confluence tools, e.g. \concon~\cite{SM14}. To this end we
need executable check functions for the syntactic properties a CTRS has to
meet in order to apply the theorem. The check functions for proper orientedness
as well as right-stability should be straightforward to implement. For
orthogonality, however, there is a small obstacle to overcome. On the one hand, in our
formalization we use the abstract notion of \emph{permutation types} 
inside the definition of conditional critical pairs,
only demanding that the set of variables is infinite. While this guarantees that
we can always rename two finite sets of variables apart, we do not directly have
an executable renaming function at our disposal. On the other hand, in the current version
of \isafor the type of variables in (standard) critical pairs is fixed to
strings, and their definition employs a concrete, executable renaming function.
Therefore, it remains to establish a suitable connection between the executable
implementation using strings and the abstract definition: for each critical pair
in the abstract definition, there is some variant that we obtain by the
executable implementation.

Moreover, Suzuki \etal~\cite{SMI95} additionally remark (without proof)
that the proof of \thmref{main} could
easily be adapted to extended proper
orientedness. To us, it is not immediately clear how to adapt our formalization.
For the time being, we leave this enhancement as future work.

\bibliographystyle{plain}
\bibliography{references}

\newcommand{\doi}[1]{
  \href{http://dx.doi.org/#1}{doi:\nolinkurl{#1}}} \newcommand{\noop}[1]{}
\begin{thebibliography}{1}

\bibitem{BN98}
Franz Baader and Tobias Nipkow.
\newblock {\em Term Rewriting and All That}.
\newblock Cambridge University Press, 1998.

\bibitem{H95}
Michael Hanus.
\newblock On extra variables in (equational) logic programming.
\newblock In {\em Proceedings of the 12th International Conference on Logic
  Programming}, pages 665--679. {MIT} Press, 1995.

\bibitem{HMS14}
Nao Hirokawa, Aart Middeldorp, and Christian Sternagel.
\newblock A new and formalized proof of abstract completion.
\newblock In {\em Proceedings of the 5th International Conference on
  Interactive Theorem Proving}, volume 8558 of {\em Lecture Notes in Computer
  Science}, pages 292--307. Springer, 2014.
\newblock \doi{10.1007/978-3-319-08970-6_19}.

\bibitem{HMS14a}
Nao Hirokawa, Aart Middeldorp, and Christian Sternagel.
\newblock Normalization equivalence of rewrite systems.
\newblock In {\em Proceedings of the 3rd International Workshop on Confluence},
  2014.

\bibitem{Isabelle}
Tobias Nipkow, Lawrence~Charles Paulson, and Makarius Wenzel.
\newblock {\em {Isabelle/HOL} - A Proof Assistant for Higher-Order Logic},
  volume 2283 of {\em Lecture Notes in Computer Science}.
\newblock Springer, 2002.
\newblock \doi{10.1007/3-540-45949-9}.

\bibitem{O02}
Enno Ohlebusch.
\newblock {\em Advanced Topics in Term Rewriting}.
\newblock Springer, 2002.

\bibitem{SM14}
Thomas Sternagel and Aart Middeldorp.
\newblock Conditional confluence (system description).
\newblock In {\em Proceedings of the Joint 25th International Conference on
  Rewriting Techniques and Applications and 12th International Conference on
  Typed Lambda Calculi and Applications}, volume 8560 of {\em Lecture Notes in
  Computer Science}, pages 456--465. Springer, 2014.
\newblock \doi{10.1007/978-3-319-08918-8_31}.

\bibitem{SMI95}
Taro Suzuki, Aart Middeldorp, and Tetsuo Ida.
\newblock Level-confluence of conditional rewrite systems with extra variables
  in right-hand sides.
\newblock In {\em Proceedings of the 6th International Conference on Rewriting
  Techniques and Applications}, volume 914 of {\em Lecture Notes in Computer
  Science}, pages 179--193. Springer, 1995.
\newblock \doi{10.1007/3-540-59200-8_56}.

\bibitem{ceta}
Ren{\'e} Thiemann and Christian Sternagel.
\newblock Certification of termination proofs using \ceta.
\newblock In {\em Proceedings of the 22nd International Conference on Theorem
  Proving in Higher Order Logics}, volume 5674 of {\em Lecture Notes in
  Computer Science}, pages 452--468. Springer, 2009.
\newblock \doi{10.1007/978-3-642-03359-9_31}.

\end{thebibliography}

\end{document}